# Photometric Redshift Estimation Using Scaled Ensemble Learning

Swagata Biswas[1], Shubhrangshu Ghosh[1], Avyarthana Ghosh[1], Yogesh Wadadekar[2], Abhishek Roy Choudhury[1],
Arijit Mukherjee[1], Shailesh Deshpande[3], and Arpan Pal[1]
[1] TCS Research, IIT Kharagpur Research Park, New Town, 700135, West Bengal, India
[2] National Centre for Radio Astrophysics, TIFR, Post Bag 3, Ganeshkhind, 411007, Pune, India
[3] TCS Research, Tata Research Development and Design Centre, 411028, Pune, India

## Abstract

The development of the state-of-the-art telescopic systems capable of performing expansive sky surveys such as the Sloan Digital Sky Survey, Euclid, and the Rubin Observatory's Legacy Survey of Space and Time (LSST) has significantly advanced efforts to refine cosmological models. These advances offer deeper insight into persistent challenges in astrophysics and our understanding of the Universe's evolution. A critical component of this progress is the reliable estimation of photometric redshifts (P$z$). To improve the precision and efficiency of such estimations, the application of machine learning (ML) techniques to large-scale astronomical datasets has become essential. This study presents a new *ensemble-based ML framework* aimed at predicting P$z$ for faint galaxies and higher redshift ranges, relying solely on optical (*grizy*) photometric data. The proposed architecture integrates several learning algorithms, including gradient boosting machine, extreme gradient boosting, $k$-nearest neighbors, and artificial neural networks, within a scaled ensemble structure. By using *bagged* input data, the ensemble approach delivers improved predictive performance compared to stand-alone models. The framework demonstrates consistent accuracy in estimating redshifts, maintaining strong performance up to $z \sim 4$. The model is validated using publicly available data from the Hyper Suprime-Cam Strategic Survey Program by the Subaru Telescope. Our results show marked improvements in the precision and reliability of P$z$ estimation. Furthermore, this approach closely adheres to—and in certain instances exceeds—the benchmarks specified in the LSST Science Requirements Document. Evaluation metrics include *catastrophic outlier*, *bias*, and *rms*.

*Unified Astronomy Thesaurus concepts:* Galaxies (573); High-redshift galaxies (734); Redshift surveys (1378); Neural networks (1933)

## 1. Introduction

The success of missions like Euclid (R. Laureijs et al. 2011; S. Ilić et al. 2022) and the Legacy Survey of Space and Time (LSST; Ž. Ivezić et al. 2019), targeting the analysis of billions of distant galaxies and quasars, depends significantly on the accuracy, precision, and efficiency of their photometric redshift (henceforth referred to as P$z$) estimation (W. A. Baum 1962; D. Koo 1985; A. J. Connolly et al. 1995; R. J. Weymann et al. 1999). P$z$s are typically derived from flux and color data across numerous filters, and their accuracy is crucial to determine the large-scale structure of the Universe, including its age and expanse. Astronomers measure the distance to objects using redshifts, then calculate the age of the farthest light that has reached us, which indicates the Universe's maximum observable distance in time. However, P$z$ estimation itself is subjected to some well-constrained boundaries set by the distance and intrinsic color of the targets. Therefore, there are significant systematic errors, due to the limited number of imaging bands available for discerning the spectral information of a galaxy (E. Jones et al. 2024, hereafter Jones2024). These errors often manifest as outliers, where the predicted P$z$ is quite different from the true redshift, increasing the bias and hence resulting in more scattered predictions (J. A. Newman & D. Gruen 2022). The situation is further complicated when the photometric errors are correlated in different bands, thereby breaking down the assumption of Gaussian noise being uncorrelated (R. Scranton et al. 2005; T. Budavári 2009).

Traditionally, P$z$ estimation has been carried out using two broad approaches: template fitting and empirical machine learning (ML). The former is informed by stellar population synthesis (SPS) models and implemented in several codes, such as Lephare (S. Arnouts et al. 1999; O. Ilbert et al. 2006), Mizuki (A. J. Nishizawa et al. 2020), and Bayesian photometric redshift (N. Benitez 2000). The traditional template-fitting approach generally suffers from inadequate spectral energy distribution (SED) templates (J. A. Newman & D. Gruen 2022), rendering it inappropriate for large-scale surveys.

On the contrary, ML techniques are based on developing a mapping from input parameters to redshift with a training set of data for which the actual spectroscopic redshifts are known, followed by applying the mappings to data for which the redshifts are to be estimated (for details, see M. Carrasco Kind & R. J. Brunner 2013; I. A. Almosallam et al. 2016; S. Cavuoti et al. 2017; M. Ntampaka et al. 2019; M. Huertas-Company & F. Lanusse 2023 and references therein). With a plethora of techniques available in ML, among the most widely used are those based on artificial neural networks (ANNs; S. Odewahn et al. 1992; A. A. Collister & O. Lahav 2004; R. D'Abrusco et al. 2007), demonstrated to estimate the P$z$ of galaxies as one of the earliest extensive use cases, as reported by L.-L. Li et al. (2007). The authors used the Sloan Digital Sky Survey (SDSS) Data Release 2 galaxy sample to use various parameters of broadband photometry, viz., magnitude, color index, and flux as inputs to compare these in their performance to estimate the redshifts. These ANN-based techniques have found applications in various other problems in astronomy like galaxy morphology classification (M. C. Storrie-Lombardi et al. 1992; M. Banerji et al. 2010; Y. Zhang et al. 2021), star–galaxy separation (S. C. Odewahn et al. 1993; N. S. Philip et al. 2002), and stellar spectral classification (R. K. Gulati et al. 1994) in various large-area astronomical sky surveys.

Other well-known and widely used techniques for estimating photometric redshifts are boosted decision trees (DTs; D. W. Gerdes et al. 2010), regression trees/random forests (RFs; L. Breiman 2001; M. Carrasco Kind & R. J. Brunner 2013), support vector machines (Y. Wadadekar 2004; E. Jones & J. Singal 2017, 2020), the direct empirical photometric method (DEmP; M. Tanaka et al. 2018), and convolutional neural networks (A. Krizhevsky 2012). Here, a huge number of galaxy images and flux measurements, captured by various galaxy survey missions, have been used as inputs, feeding flux, color, and other information for deriving the photo-$z$ estimates.

Another kind of widely used neural network technique, the probabilistic Bayesian neural network (BNN; D. F. Specht 1990; A. Filos et al. 2019; M. W. Dusenberry et al. 2020), has proved useful, providing better uncertainty representations, point predictions, and interpretability. However, these approaches depend strongly on the availability of large, high-quality training datasets. This proves quite challenging, especially for galaxies at high redshifts, due to the inherent faint nature of such galaxies. Also, the disproportionately low fraction of high-redshift galaxies ($z > 2.5$) with accurate spectroscopic redshifts as compared to the low-redshift ones (M. Wyatt & J. Singal 2021; Jones2024) poses a challenge for training. Therefore, photometric redshift determination of high-redshift galaxies is not well explored in the literature. In fact, this holds true for most of the existing models that have been implemented for the Rubin/LSST (up to $z = 3$; M. Tanaka et al. 2018; S. Schmidt et al. 2020; S. Schuldt et al. 2021), which aims at observing ∼20 billion galaxies, ∼17 billion resolved stars, and ∼6 million orbits of solar system bodies[4] (for more details, refer to the LSST Science Requirements Document[5]). Therefore, accurately estimating redshifts for high-$z$ galaxies remains a challenging task, necessitating the development of a sophisticated algorithm capable of delivering precise and reliable predictions.

To bridge this gap, we introduce a novel scaled ensemble-based framework for photometric redshift estimation. There are several AI/ ML/ DL models like RF, BNN, etc. for regression problems. Each of these models has its own advantages and limitations. In this work, we have adopted the ensemble-based approach, where we can exploit the benefits of using multiple models. We have extensively explored multiple ML algorithms (like gradient boosting machine, GBM; random forest, RF; extreme gradient boosting, XGB; CatBoost; linear regression; $k$-nearest neighbor, KNN; extremely randomized trees, and ANN) in different combinations to find suitable candidates to build the first layer of our architecture. Based on their performance and following extensive experimentation, we have optimally integrated four learning models—viz., GBM (J. H. Friedman 2001), XGB (C. Wade & K. Glynn 2020), KNN (T. Cover & P. Hart 1967), and ANN (A. A. Collister & O. Lahav 2004)—within a scalable ensemble structure, utilizing bagged (L. Breiman 1996) input to enhance predictive performance beyond that of individual learners. Our proposed scaled ensemble architecture exploits the diversity of the four base learners (GBM, XGB, KNN, and ANN) by applying bagging to the input data, thereby improving generalization and robustness. The first layer captures diverse decision boundaries, combining the interpretability of tree-based methods with the deep feature extraction capabilities of neural networks. A weighted ensemble in the second layer effectively balances the contributions of these models, minimizing bias and variance while enhancing predictive performance. This hierarchical design reduces overfitting and improves adaptability to complex, high-dimensional data distributions, particularly for photometric redshift estimation in the higher, more challenging redshift range. An ablation study was conducted to evaluate the actual impact of these design choices, confirming that the proposed ensemble consistently outperformed its individual base models. The primary aim of this study was to introduce a novel architecture capable of achieving superior performance in photometric redshift estimation. Comparative analysis with the latest state-of-the-art methods further validates the effectiveness of our approach, particularly for the upper and more demanding ranges of $z$. This methodology enables accurate redshift estimation even at higher redshift ranges, specifically within $0 < z < 3.99$. Our experimental results demonstrate a substantial improvement in P$z$ estimation up to $z = 3.99$, leveraging data from the Hyper Suprime-Cam (HSC) hosted by the Subaru Telescope. Furthermore, our model closely aligns with the specifications outlined in the LSST Science Requirements Document. The effectiveness of the proposed method is assessed using key evaluation metrics, including *catastrophic outlier*, *bias*, and *rms*.

The remainder of this paper is organized as follows. Section 2 provides a short discussion and comparative analysis of existing P$z$ estimation techniques proposed for the Rubin/ LSST program. Section 3 details the proposed algorithm and dataset, followed by a presentation of results and analysis in Section 4. Finally, conclusions are drawn in Section 5.

## 2. Existing Photometric Redshift Estimators for Rubin/LSST

The Rubin Observatory data management (DM) team received 20 submissions to the P$z$ letters of recommendation (LOR) process (M. Graham et al. 2023). Among these, there were 19 LOR and 1 non-LOR that describe the Dark Energy Science Collaboration photo-$z$ activities. The 19 LOR received included 12 specific algorithms for P$z$ estimation, 6 that represented scientific use cases, and 1 that was dedicated to future software development using deep probabilistic networks.

The six LORs representing scientific use cases include one solution focused on galaxies, two solutions focused on dark energy, and three solutions focused on active galactic nuclei (AGN).

The 12 LORs that proposed specific algorithms include the following.

1. Four ML-based algorithms, namely, GPz (Z. Gomes et al. 2018), DEmP (B. Hsieh & H. Yee 2014), PZFlow (J. F. Crenshaw et al. 2024), and DNF (J. De Vicente et al. 2016).

---

[4] https://rubinobservatory.org/for-scientists/rubin-101/key-numbers
[5] https://docushare.lsstcorp.org/docushare/dsweb/Get/LPM-17

 2. Three template-fitting-based algorithms, namely, LePhare[6,7], Phosphorous (M. Tucci et al. 2025), and BPZ (N. Benitez 2000).
 3. Two hybrid algorithms, incorporating both ML and template-fitting techniques, namely, Delight (B. Leistedt & D. W. Hogg 2017) and ML-accelerated hierarchical SPS[8] models.
 4. Three algorithms aiming at postprocessing to enhance P$z$ estimates, for example, combining P$z$ estimates, recalibrating probability density functions (pdfs), refining outlier flags, etc.

GP$z$, DEmP, DNF, LePhare, and BPZ claimed in the LOR that the respective software is ready. Also, they can meet the various scientific, performance, and technical requirements as described in the call. However, three LORs—PZFlow, Delight, and Phosphorous—were still in development and did not meet all the necessary criteria at the time the LORs were submitted.

Eventually, eight P$z$ estimators were recognized formally for inclusion in the LSST objects catalog. They are GP$z$, DEmP, DNF, LePhare, BPZ, PZFlow, Delight, and Phosphorous as described in Table 1. DM will prioritize the implementation and validation of more established estimators due to its budgeted resources. Therefore, GP$z$, DEmP, DNF, LePhare, and BPZ would be preferred for testing and validation first.

The ML-accelerated hierarchical SPS models have yet to prove whether they will be able to meet the requirements, but they seem quite promising. Other promising P$z$ estimators as mentioned in S. Schmidt et al. (2020) for LSST objects are ANNz2 (I. Sadeh et al. 2016), EAZY (G. B. Brammer et al. 2008), FlexZ-Boost[9] (N. Dalmasso et al. 2020), METAPhoR (S. Cavuoti et al. 2017), SkyNet (T. Nakajima et al. 2020), and TPZ (M. Carrasco Kind & R. J. Brunner 2013).

## 3. Data and Methods

### 3.1. Dataset Used

Our primary goal is to make photometric redshift predictions for the upcoming LSST to be carried out by the Vera Rubin Observatory. The commencement of the 10 yr program of the LSST is expected by the end of 2025. Jones2024, aiming to develop an algorithm for photometric redshift estimation for the LSST, are currently using the Subaru HSC Strategic Survey Program (SSP; H. Aihara et al. 2018) dataset for algorithm development, validation, and testing. The HSC-SSP dataset is considered to be the most representative present-day counterpart for LSST, given its multifilter coverage in the optical band over a fairly large area of sky. Based on HSC-SSP data, Jones2024 compiled a catalog of galaxies (E. Jones et al. 2021) with similar optical photometry and depth to the LSST survey. Their catalog has 22 columns. It comprises data on 286,401 galaxies. We divided this dataset into three subsets. The first subset comprises galaxies with $0 < z \leqslant 0.5$. This subset is chosen for easy comparison with the P$z$ obtained using other surveys, like SDSS. The second subset comprises data points with $0 < z \leqslant 1.5$. This subset is chosen as it consists of a reasonably large number of galaxies. Also, it allows for comparison with Jones2024, who study galaxies in the same redshift range. In the third subset, we consider data points with $0 < z \leqslant 3.99$. Here, we select the complete dataset for testing our proposed model. For each galaxy, the spectroscopic redshift, the error in observing the spectroscopic redshift, the flux intensity in each of the five *grizy* bands, and the error in observing the flux intensity for each band are used in our analysis. Including the photometry data in infrared bands would likely improve the photo-$z$ estimation. However, since the LSST survey will provide observations only in the optical bands (300−1100 nm; J. F. Crenshaw et al. 2025), we have chosen to use the optical bands only for our analysis.

### 3.2. Network Architecture

We present a novel, scalable, ensemble-based framework for photometric redshift estimation that integrates multiple ML models to achieve enhanced predictive performance. The framework combines GBM, XGB, KNN, and ANN within a unified ensemble architecture. By applying bagging to the input data, the ensemble leverages the diversity of these base learners to improve generalization and reduce overfitting. The architecture is structured in two layers. The initial layer incorporates GBM, XGB, KNN, and ANN to capture diverse decision boundaries, effectively blending the explainability of tree-based methods with the expressive capability of neural networks. The second layer forms a weighted ensemble of the first-layer outputs, effectively balancing the strengths of each model to minimize bias and variance. This hierarchical design not only enhances robustness but also improves adaptability to complex, high-dimensional data distributions, particularly in the challenging high-redshift regions. The proposed network architecture is shown in Figure 1. The input to the proposed network comprises the flux measured in five optical bands (*grizy*). The network architecture has two layers. In the first layer, we have used the GBM, XGB, KNN, and ANN models. For each of these models, the input to the model is bagged and then fed to the model. In the second layer or output layer of the network, we have used the scaled ensemble technique. These models are described briefly below.

*Bagging.* Bagging (L. Breiman 1996), also known as the bootstrap aggregating method, is an ML ensemble learning technique. In this technique, the input dataset ($X$) is divided into, say, $p$ subsets, namely, $X_1, X_2, ..., X_i, ..., X_p$ as shown in Equation (1). Each subset, $X_i$, is then used to train a base learner model ($M_i$) independently. The output of all the base learner models ($M_1, M_2, ..., M_i, ..., M_p$) is aggregated (in case of regression) or chosen using a voting technique (in case of classification), and the final output is predicted. This technique helps in introducing diversity among the models. It also aids in canceling out the effect of the errors and increasing the reliability and stability in the prediction.

The complete dataset is split in an 80:10:10 ratio representing the training, validation, and testing dataset, respectively, following Jones2024. The training dataset is further subdivided into $p$ subdatasets. Here, $p$ is considered to be 10. Therefore, the training dataset ($X$) is split into 10 subdatasets ($X_1, X_2, ..., X_i, ..., X_{10}$). Now, each dataset $X_i$ is used to train the GBM, XGB, KNN, and ANN models. The validation dataset, i.e., 10% of the complete dataset, and the testing dataset, i.e., 10% of the complete dataset, remain

---

[6] https://lephare.readthedocs.io/en/latest/
[7] https://community.lsst.org/t/lor-for-the-lephare-pz-estimator/5879
[8] community.lsst.org/t/lor-ml-accelerated-hierarchical-stellar-population-synthesis-sps-models/5875
[9] https://github.com/LSSTDESC/rail_flexzboost/

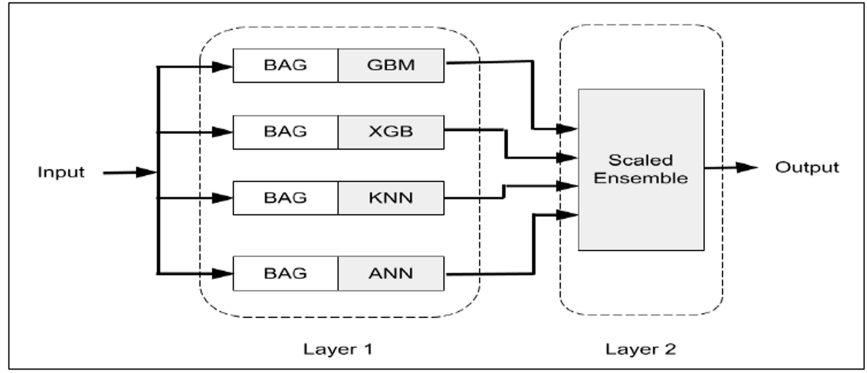

**Figure 1.** Network architecture for our scaled ensemble model. Abbreviations used in the figure: bagging (BAG), GBM, XGB, KNN, ANN.

the same for all these models:

$$\text{Bag}(X, Y, p) = \{(X_1, Y_1), (X_2, Y_2), ..., (X_i, Y_i), ..., (X_p, Y_p)\}. \quad (1)$$

*GBM.* GBM (J. H. Friedman 2001) is a type of ensemble ML algorithm. It follows a sequential training process. Let us consider that the training set is represented as $(X, Y)$, where $X$ represents the input and $Y$ represents the labeled output. Now, let us assume that there are $n$ models $\{M_1^G, M_2^G, ..., M_j^G, ..., M_n^G\}$ as shown in Equation (2). In the first step, $(X, Y)$ is fed as input to model $M_1^G$. Let us consider that the predicted output set is $Y_1$. The error in prediction in the first step is evaluated as $E_1 = Y - Y_1$. In the next step, the input to the next model $M_2^G$ is given as $(X, E_1)$. This helps model $M_2^G$ to rectify the error in the prediction made in the previous model, i.e., $M_1^G$. This process is continued until the last model, i.e., $M_n^G$. Therefore, we can see that at each step, any model $M_j^G$ tries to resolve the errors made in previous models by taking the error in prediction as input. Hence, we can say that by using this method, we can build a stronger model from a set of weaker models, improving the predictive accuracy.

Here, we have used the gradient boosting regressor model from the scikit-learn library in Python. The number of DTs is initialized to 100, and the learning rate is considered to be 0.1. Using each $X_i$, we get a model $M_i^G$ as shown in Equation (3). Finally, the best model is chosen for which the loss is minimum:

$$M^G = \{M_1^G, M_2^G, ..., M_j^G, ..., M_n^G\}, \quad (2)$$

$$\text{GBM}(X_i, Y_i, M^G) = M_i^G. \quad (3)$$

*XGB.* XGBoost (C. Wade & K. Glynn 2020) is a type of ensemble learning. The goal of this algorithm is to combine a series of weak learner models to develop a strong model. This is done by minimizing the loss function iteratively using gradient descent method. This approach is similar to GBM. However, XGB has few added benefits. XGBoost adopts a regularization technique to avoid overfitting. It has the ability to speed up the tree-building process through parallelization. XGBoost is capable of handling missing data. Unlike GBM, XGBoost is not limited to local optimization. In GBM, the split with negative loss is always pruned. However, in XGBoost, the split with negative loss may be retained if the overall split is positive or the set value in the max depth parameter is not yet reached.

Here, we have used the XGBoost regressor model from the XGBoost package in Python. The number of DTs is initialized to 10. Therefore, the value of $n$ in Equation (4) is 10. Similar to GBM model training, using each dataset $X_i$ formed using the bagging technique, we obtain a model $M_i^X$ as shown in Equation (5). The model with minimum loss is selected:

$$M^X = \{M_1^X, M_2^X, ..., M_j^X, ..., M_n^X\}, \quad (4)$$

$$\text{XGB}(X_i, Y_i, M^X) = M_i^X. \quad (5)$$

*KNN.* KNN (T. Cover & P. Hart 1967; K. J. Luken et al. 2019) is a type of supervised ML algorithm. Here, the predicted value of the target point is evaluated by taking the average of the values of $k$-nearest neighbors (in case of regression) or by voting (in case of classification).

Here, we have used the KNN regressor model from the scikit-learn library in Python. We have assumed the value of $k$ to be 5. Using each dataset $X_i$, obtained using the bagging technique, we obtain a model $M_i^K$ as shown in Equation (7). The model with minimum loss is chosen:

$$M^K = \{M_1^K, M_2^K, ..., M_j^K, ..., M_n^K\}, \quad (6)$$

$$\text{KNN}(X_i, Y_i, M^K) = M_i^K. \quad (7)$$

*ANN.* Here, we have used the Sequential class in the Keras (A. Géron 2022) library to build the ANN model (A. A. Collister & O. Lahav 2004). The ANN model has four layers comprising 1000, 500, 250, and 1 neuron in each layer sequentially. For the first three layers, the activation function used is *relu*, and for the last layer, the activation function used is *linear*. The optimizer used is *RMSProp*. We have used the early stopping callback feature. The objective is to minimize the loss. The patience value used is 50, the number of epochs is 50, and the batch size is 50:

$$M^A = \{M_1^A, M_2^A, ..., M_j^A, ..., M_n^A\}, \quad (8)$$

$$\text{ANN}(X_i, Y_i, M^A) = M_i^A. \quad (9)$$

*Scaled ensemble.* The second layer of our proposed model comprises the scaled ensemble. Here, the predicted output from all four models in the first layer, i.e., $y_{\text{pred}}^G$ (GBM), $y_{\text{pred}}^X$ (XGB), $y_{\text{pred}}^K$ (KNN), and $y_{\text{pred}}^A$ (ANN), is aggregated to the final predicted value, $y_{\text{pred}}$, using the scaled ensemble as shown in Equations (12) and (13). The weights given to each predicted

Table 1
Pz Estimators Whose Outputs Will Be Included in the LSST Object Catalog

| Pz Estimator | Approach | Novelty | Gap |
| --- | --- | --- | --- |
| GPz (Z. Gomes et al. 2018) | Gaussian process regression | Included near-IR filters, angular size as features | Target galaxies at $z < 0.5$ |
| DEmP (B. Hsieh & H. Yee 2014) | Empirical method | Resolves the issue of suitability of the empirical function used by choosing only the nearest neighbors in the multidimensional color–magnitude space for each galaxy to derive the Pz for that galaxy | The training set needs to be robust in all aspects under consideration, in fact, better than the target set; therefore, lack of information, even for a small percentage of objects being studied, leads to significantly discrepant results |
| DNF (J. De Vicente et al. 2016) | Linear combination of multiband fluxes | Best approximation for photometric redshift estimation, capable of yielding Pz estimates with very low dispersion (similar to ANNs), unlike the relatively larger dispersion values from the KNN technique, i.e., among the other nearest-neighbor approaches explored | Designed only for galaxies |
| LePhare[a,b] | Template fitting | Involves fitting of SEDs to a dataset of photometric fluxes or apparent magnitudes as well as stellar templates to compute photometric redshifts for galaxies and AGN; also implemented for deriving galaxy physical parameters; can reliably compute photo-z for AGN and QSO | Suffers from significant inaccuracies for computing photometric redshifts of faint or high-redshift objects, albeit bias in training datasets; susceptible to SED template quality and dust attenuation of the observations used and faces difficulty in generating valid pdfs of redshift, which are crucial for characterizing the accuracy of the redshift estimate |
| BPZ (N. Benitez 2000) | Template fitting | By selecting appropriate threshold values, outliers can be effectively removed from the sample set, leading to more accurate error estimates compared to many other ML techniques | Choice of the optimized threshold value holds significance for the correct estimation |
| PZFlow (J. F. Crenshaw et al. 2024) | Hybrid (template fitting and ML) | Enables forward modeling of galaxy catalogs | Exhibits mode-covering behavior, i.e., tends to give broad posteriors, which can have a negative impact for large training datasets |
| Delight (B. Leistedt & D. W. Hogg 2017) | Combination of template fitting with ML-based techniques | Training data are a combination of spectroscopic and photometric data acquired over multiple bands with reliable redshifts, with not much dependence on the spatial overlap with the target survey of interest or even involving the same photometric bands | Under development; need to meet all evaluation criteria |
| Phosphorous (M. Tucci et al. 2025) | Template fitting | Deployed full Bayesian framework with flexible prior parameters and sampling from multidimensional and marginalized posteriors | Under development; need to meet all evaluation criteria |

**Notes.**
[a] https://lephare.readthedocs.io/en/latest/
[b] https://community.lsst.org/t/lor-for-the-lephare-pz-estimator/5879

value are $w^G$ ($y_{\text{pred}}^G$), $w^X$ ($y_{\text{pred}}^X$), $w^K$ ($y_{\text{pred}}^K$), and $w^A$ ($y_{\text{pred}}^A$). The model with minimum loss is given the highest weightage as depicted in Equation (14). The loss is represented as $loss^G$ ($M^G$), $loss^X$ ($M^X$), $loss^K$ ($M^K$), and $loss^A$ ($M^A$). With increasing loss value, the weightage given decreases proportionally:

$$M^Q(X_{\text{test}}) = y_{\text{pred}}^Q, \quad (10)$$

$$Q = \{G, X, K, A\}, \quad (11)$$

$$M^E(y_{\text{pred}}^G, y_{\text{pred}}^X, y_{\text{pred}}^K, y_{\text{pred}}^A) = y_{\text{pred}}, \quad (12)$$

$$y_{\text{pred}} = w^G * y_{\text{pred}}^G + w^X * y_{\text{pred}}^X + w^K * y_{\text{pred}}^K + w^A * y_{\text{pred}}^A, \quad (13)$$

$$w^Q \propto \frac{1}{\text{loss}^Q}, \quad (14)$$

$$w^G + w^X + w^K + w^A = 1. \quad (15)$$

The proposed scaling ensemble architecture leverages the diversity of multiple base learners—GBM, XGB, KNN, and ANN—by bagging the input data, enhancing generalization and robustness. The first layer captures varied decision boundaries, benefiting from both tree-based models' interpretability and neural networks' deep feature learning. The second layer, a weighted ensemble, optimally balances these models' contributions, reducing bias and variance while improving predictive accuracy. This hierarchical structure mitigates overfitting and enhances adaptability to complex, high-dimensional data distributions for photometric redshift estimation in the upper and more demanding range of $z$.

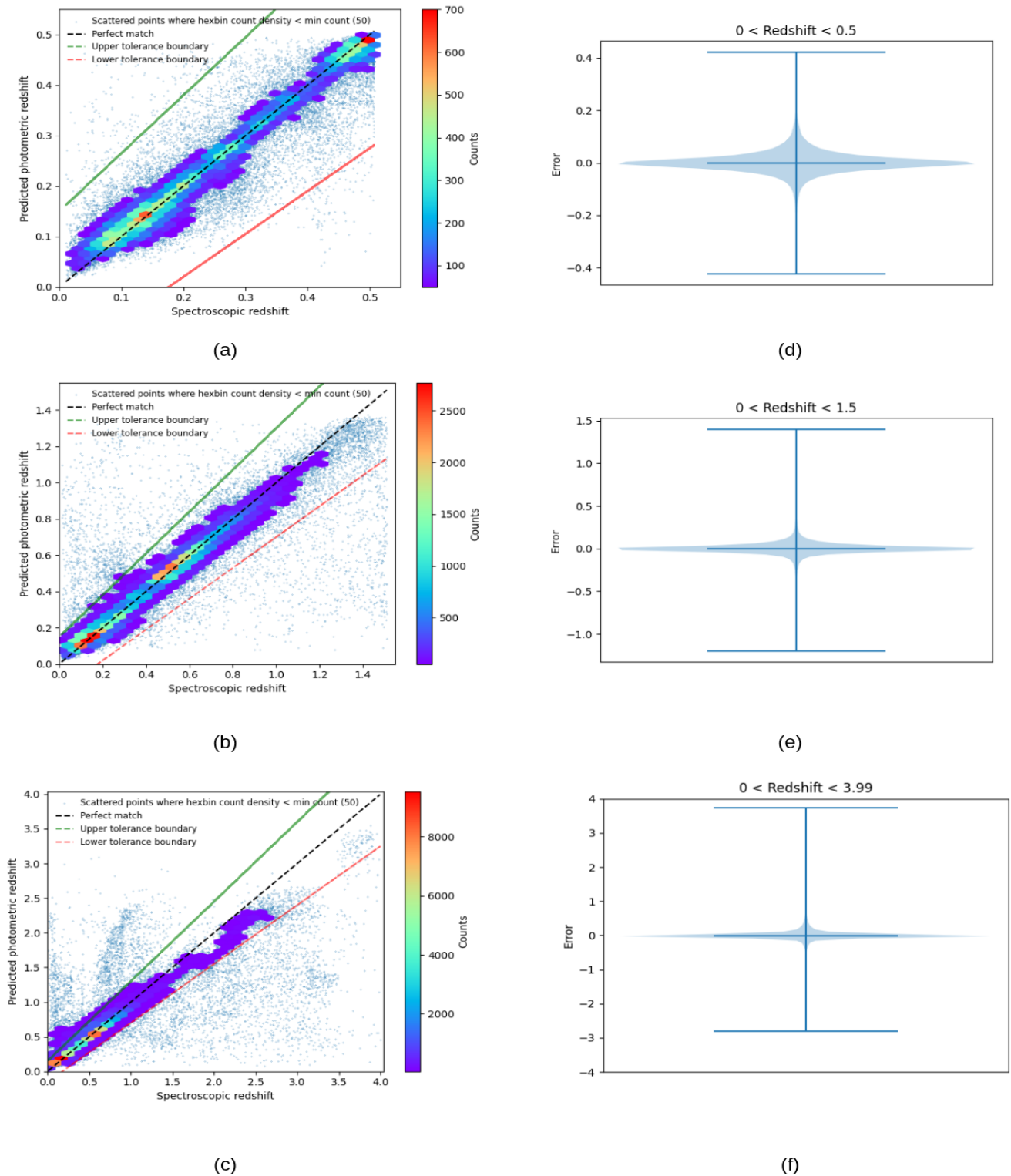

**Figure 2.** Hexbin plot for (a) $0 < z < 0.5$, (b) $0 < z < 1.5$, and (c) $0 < z < 3.99$ with a minimum bin density count of 50. The scatter plot is embedded to represent points lying below the minimum bin density count. The 45° line represents the perfect match. The upper tolerance boundary can be represented as $y_{\text{pred}} = 1.15 y_{\text{test}} + 0.15$, and the lower tolerance boundary can be represented as $y_{\text{pred}} = 0.85 y_{\text{test}} - 0.15$, where $y_{\text{pred}}$ represents the predicted photometric redshift and $y_{\text{test}}$ represents the actual spectroscopic redshift. Points lying outside the tolerance lines are outliers, by our definition. Violin plots for (d) $0 < z < 0.5$, (e) $0 < z < 1.5$, and (f) $0 < z < 3.99$ depict the distribution of the error in prediction. The errors are predominantly centered around 0, with a narrow and symmetrical shape. The tight clustering around 0 suggests that most predictions are close to the actual values. The slim shape of the violin plots suggests low error margins, while their symmetry indicates a lack of systematic bias, signifying that the model consistently neither overestimates nor underestimates the actual values.

**Table 2**
Comparison with Jones2024 Results

| Network | Redshift Range | Catastrophic Outlier | rms | Bias |
|---|---|---|---|---|
| Mizuki | ⩽1.5 | 0.102 | 0.307 | 0.011 |
| DEmP | ⩽1.5 | 0.092 | 0.277 | 0.003 |
| XGBoost | ⩽1.5 | 0.022 | 0.149 | 0.002 |
| SPIDERz | ⩽1.5 | 0.051 | 0.199 | 0.002 |
| RF | ⩽1.5 | 0.006 | 0.088 | 0.001 |
| BNN | ⩽1.5 | 0.023 | 0.174 | 0.013 |
| Our ($z \leqslant 0.5$) | ⩽0.5 | 0 | 0.03 | 0.0000294 |
| Our ($z \leqslant 1.5$) | ⩽1.5 | 0 | 0.07 | 0.0000295 |
| Our ($z \leqslant 3.99$) | ⩽3.99 | 0.0224 | 0.13 | 0.0127 |

### 3.3. Experimental Setup

*Dataset preprocessing.* All galaxies having a flux error greater than 10% are removed. The dataset is then divided into three sets by binning in redshift. The first set comprises 44,923 galaxies with redshift ($z$) less than or equal to 0.5. The second set comprises 78,660 galaxies with $0 \leqslant z \leqslant 1.5$. The third set comprises 84,446 galaxies with $0 < z < 3.99$. Initially, we tested our model in a lower redshift range ($0 \leqslant z \leqslant 0.5$) where the signal-to-noise ratio is typically high. Then we have incrementally increased the redshift range ($0 \leqslant z \leqslant 1.5$) to compare the performance of our proposed model to the Jones2024 model. Finally, we have tested our model extensively at higher redshift ranges ($0 \leqslant z \leqslant 3.99$) to study the effect of decreased mean signal-to-noise ratio on the performance metrics studied. In our experiment, we have used the flux intensity in the five bands, i.e., *grizy*, as the feature set. The spectroscopic redshift is considered to be the ground truth. We split the dataset into training, testing, and validation set as per Jones2024. Of the complete dataset, 80% is chosen as the training set, 10% is chosen as the validation set, and the remaining 10% is chosen as the test set. The validation set and the test set remain the same for each algorithm in all layers in the network architecture. The hyperparameter tuning is done using the validation set. The hyperparameters are chosen using the grid search method. At first, we defined the hyperparameter grid specifying the range of discrete values for each of the hyperparameters. In our model, we have used default hyperparameter values in some cases, and a few are obtained after hyperparameter tuning. The grid search method generates all possible combinations of each of the hyperparameters. For each unique combination of hyperparameters generated, a model is trained on the training dataset. The performance of this trained model is then evaluated on the validation dataset using the loss function. This evaluation provides a performance score for each hyperparameter combination. After evaluating all combinations, the set of hyperparameters that yielded the best performance score, i.e., the minimum loss, on the validation dataset is selected as the optimal hyperparameter set. The final model is then trained on the entire training dataset using these chosen optimal hyperparameters. This model is expected to generalize well to unseen data. The feature set and the ground truth are fed to the first layer of our proposed model. As shown in Figure 1, in the first layer, the input data are first bagged and then fed to the GBM, XGB, KNN, and ANN models in parallel.

*Loss function.* Here, we have used a custom loss function. Let us consider the actual redshift to be $y_{\text{test}}$ and the predicted redshift value to be $y_{\text{pred}}$. Therefore, the loss can be evaluated as given in Equation (17):

$$\Delta z = |y_{\text{pred}} - y_{\text{test}}|, \quad (16)$$

$$\text{loss} = \frac{\Delta z}{1 + \Delta z}. \quad (17)$$

Here, training directly on $\Delta z$ makes the loss dominated by high-$z$ objects (where $z$ is larger), thereby biasing the model. Using $\Delta z/1 + \Delta z$ gives roughly uniform weight across redshifts and prevents the model from overfitting the high-$z$ tail.

*Performance metrics.* According to the LSST Science Requirements Document, the four significant performance metrics are outlier ($O$), bias ($b$), catastrophic outlier ($O_c$), and rms. The performance metrics are evaluated as described in Equations (18), (19), (20), and (21):

$$O = \frac{|y_{\text{pred}} - y_{\text{test}}|}{1 + y_{\text{test}}}, \quad (18)$$

$$O_c = |y_{\text{pred}} - y_{\text{test}}|, \quad (19)$$

$$b = \frac{y_{\text{pred}} - y_{\text{test}}}{1 + y_{\text{test}}}, \quad (20)$$

$$\text{rms} = \sqrt{\frac{1}{n_{\text{gal}}} \sum \left( \frac{y_{\text{pred}} - y_{\text{test}}}{1 + y_{\text{test}}} \right)^2}. \quad (21)$$

An astronomical object is considered to be an outlier if $O > 0.15$ and a catastrophic outlier if $O_c > 1.0$. According to the LSST requirement, rms < 0.02, $b < 0.003$, and $O_c < 10\%$ of the total sample size.

*System configuration.* We used an A5500 GPU for network model training and evaluation. The average training time required was about 21,600 s. The average individual model training time is in the order ANN > GBM > XGB > KNN ranging between 19,800 and 25.2 s. The average evaluation time for testing the set using the scaled ensemble model is approximately 4.4 s. The average evaluation time of the testing set using individual models is in the order KNN > ANN > GBM > XGB ranging between 4 and 0.15 s.

### 4. Results and Discussion

We have performed three experiments. In the first experiment, we have considered data points where $0 < z < 0.5$. We have trained and tested using our proposed model as explained in Section 3.2. We have visualized the true redshift and the predicted redshift using the hexbin plot and the violin plot as shown in Figure 2. In the hexbin plot, we have considered the minimum bin density to be 50. The 45° line represents the perfect match. We have considered the upper tolerance

Table 3
Ablation Study Performance Evaluation of Individual Models (ANN, GBM, XGB, and KNN) and the Scaled Ensemble Model for Unbagged and Bagged Datasets

| Performance | Unbagged | | | | | Bagged | | | | | $\Delta^{\%}_{I_{UB}-I_B}$ |
|---|---|---|---|---|---|---|---|---|---|---|---|
| | ANN | GBM | XGB | KNN | Ensemble | ANN | GBM | XGB | KNN | Ensemble | |
| Metrics | | | | | ($I_{UB}$) | | | | | ($I_B$) | |
| MAE | 0.22953 | 0.27954 | 0.06566 | 0.02338 | 0.04659 | 0.22 | 0.26703 | 0.06371 | 0.02279 | 0.0451 | 3.1981 |
| rms | 0.26803 | 0.28869 | 0.14143 | 0.12781 | 0.13618 | 0.25994 | 0.28083 | 0.13499 | 0.12217 | 0.13 | 4.5381 |
| Bias | 0.03897 | 0.0454 | 0.00001 | 0.02585 | 0.01341 | 0.03748 | 0.0437 | 0.00001 | 0.02465 | 0.0127 | 5.2946 |
| $O_c$ | 0.04584 | 0.05292 | 0.00749 | 0.03855 | 0.02349 | 0.04392 | 0.05015 | 0.00716 | 0.03693 | 0.0224 | 4.6403 |

**Note.** Performance metrics evaluated: MAE, rms error, bias, and catastrophic outlier ($O_c$). Percentage improvement in performance by using bagging is represented as $\Delta^{\%}_{I_{UB}-I_B}$, where $I_{UB}$ represents the performance score for unbagged data and $I_B$ represents the performance score for bagged data.

boundary and the lower tolerance boundary to be represented using Equations (22) and (23):

$$y_{\text{pred}} = 1.15 y_{\text{test}} + 0.15, \quad (22)$$

$$y_{\text{pred}} = 0.85 y_{\text{test}} - 0.15. \quad (23)$$

In Figures 2(a)–(c), where the density of data points is below the minimum density count of the hexbin, the data points are visualized using a scatter plot. It is observed that the maximum data points lie within the hexbins and are in close proximity to the perfect match line. Other data points plotted using the scatter plot mostly lie within the tolerance boundaries. Few data points lie outside the tolerance boundary. Therefore, we can conclude that the predictive accuracy is consistent throughout the selected redshift range, i.e., $0 < z < 0.5$. We have also used the violin plot to analyze our results. It is used to visualize the error in prediction. It is observed that the data points lie close to 0. Therefore, we can conclude that the prediction error is minimal.

We have performed the same experiment using data points where $z < 1.5$. The results are represented using the hexbin plot (Figure 2(b)) and the violin plot (Figure 2(e)). The tolerance boundaries and the minimum bin density count remain the same as considered for the first experiment. If we compare the hexbin plot for $z < 0.5$ (Figure 2(a)) and $z < 1.5$ (Figure 2(b)), it is observed that the number of data points is more unequally distributed. This is because there are more galaxies available at lower redshift. Here also, most data points lie within the defined tolerance boundaries. Very few data points lie beyond the tolerance boundaries. Therefore, we can infer that the predictive accuracy is consistent throughout the given redshift range. Also, the violin plot (Figure 2(e)) indicates a relatively small error in prediction.

Next, we repeated the same experiment using data points where $z < 3.99$. The results are visualized in Figures 2(c) and (f). It is observed that in this case too, the distribution in redshift is biased toward lower redshift as compared to $z < 0.5$ (Figure 2(a)) and $z < 1.5$ (Figure 2(b)). The error in prediction, as visualized using the violin plot (Figure 2(f)), also lies predominantly near 0, indicating a minor error in prediction.

We next evaluated the performance metrics for all three datasets ($z < 0.5$, $z < 1.5$, $z < 3.99$) as explained in Section 3.3. The results are tabulated in Table 2. The Jones2024 study reveals that by using RF and BNN, the researchers have achieved the best results when compared to other solutions like Mizuki, DEmP, XGBoost, and SPIDERz. However, they have studied the performance metrics for the $0 < z < 1.5$ range only. In our experiment, we have extended our study to the $0 < z < 3.99$ range. The results clearly depict that using our model, we are able to meet or closely align with the requirements as stated in the LSST Science Requirements Document. Also, the catastrophic outlier, rms error, and bias are significantly smaller using our proposed model as compared to the RF and BNN models of Jones2024.

*Ablation study.* Here, we have studied the performance of each individual model as well as the ensemble model. We have also studied the effect of bagged and unbagged data on the performance metrics. The performance metrics observed here are mean average error (MAE), rms, bias, and catastrophic outlier. The MAE or the loss is evaluated using Equation (17). The other performance metrics are evaluated using Equations (18)–(21). The results are depicted in Table 3. Here, we have presented the results for $0 < z < 3.99$. The weight given to each model is dependent on the MAE score gained using the individual model. In the ensemble model, the sum of the weights given to all models equals 1. It is observed that the performance improvement evaluated using any of the abovementioned performance metrics, by using the bagged dataset, ranges between 3% and 6%. This ablation study enables us to understand the benefit gained by integrating these individual models.

The scaling ensemble architecture capitalizes on the complementary strengths of diverse base learners—GBM, XGB, KNN, and ANN—by applying bagging to the input data, thereby improving model robustness and generalization. In the first layer, each model captures distinct decision boundaries, combining the interpretability of tree-based approaches with the deep feature extraction capabilities of neural networks. By employing a bagged input strategy, our method enhances predictive accuracy, surpassing the performance of individual learners. The second layer employs a weighted ensemble to effectively integrate these outputs, striking a balance that minimizes both bias and variance while enhancing predictive accuracy. This hierarchical design reduces overfitting and offers improved adaptability to complex, high-dimensional data structures, particularly for photometric redshift estimation in the higher and more challenging $z$ range.

## 5. Conclusion

Estimating redshifts for large-scale structures, such as galaxies and their clusters, plays a crucial role in constraining the evolution of our Universe (V. R. Eke et al. 1996; N. A. Bahcall et al. 1997; J. P. Henry 2000, among others) and provides insights into the nature of the pervasive dark matter and dark energy. However, current state-of-the-art methodologies face significant challenges, particularly due to the incompleteness of spectroscopic training samples, which serve as ground truth for redshift estimation. This limitation is even more pronounced for high-$z$ samples. Additionally, at higher redshifts, the signal strength diminishes, and this poorer signal-to-noise ratio leads to greater difficulties in spectroscopic redshift estimation.

To mitigate these limitations, we introduce an innovative scaled ensemble framework for photometric redshift estimation utilizing *grizy* photometry data. This approach integrates a diverse set of ML models—including GBM, XGB, KNN, and ANN—within a structured ensemble architecture. By employing a bagged input strategy, our method enhances predictive accuracy, surpassing the performance of individual learners.

This framework enables precise photometric redshift estimation even at higher redshifts, specifically within $0 < z < 3.99$. While most prior studies have primarily focused on lower redshift ranges ($0 < z < 1.5$), our model achieves comparable accuracy across an extended range of redshift. Furthermore, it closely aligns with the specifications outlined in the LSST Science Requirements Document and outperforms existing methods, as reported in Jones2024. The model's effectiveness is evaluated using major performance metrics, including catastrophic outlier (also see M. Banerji et al. 2008), bias, and rms.

For future work, we intend to validate our model using the upcoming LSST dataset and expand our evaluation by incorporating additional performance metrics, such as *scatter*, *loss*, and *coverage*. Given the computational demands of large-scale surveys, optimizing our approach for minimal time, memory, and processing complexity is also a priority. In subsequent studies, we aim to refine our model further to enhance efficiency and overall performance.

## Acknowledgments

We thank the anonymous referee for the insightful comments and suggestions that have helped improve both the content and presentation of this paper. Y.W. acknowledges the support of the Department of Atomic Energy, Government of India, under project No. 12-R&D-TFR5.02-0700.

The Hyper Suprime-Cam (HSC) collaboration includes the astronomical communities of Japan and Taiwan and Princeton University. The HSC instrumentation and software were developed by the National Astronomical Observatory of Japan (NAOJ), the Kavli Institute for the Physics and Mathematics of the Universe (Kavli IPMU), the University of Tokyo, the High Energy Accelerator Research Organization (KEK), the Academia Sinica Institute for Astronomy and Astrophysics in Taiwan (ASIAA), and Princeton University. Funding was contributed by the FIRST program from the Japanese Cabinet Office, the Ministry of Education, Culture, Sports, Science and Technology (MEXT), the Japan Society for the Promotion of Science (JSPS), Japan Science and Technology Agency (JST), the Toray Science Foundation, NAOJ, Kavli IPMU, KEK, ASIAA, and Princeton University.

This paper makes use of software developed for the Large Synoptic Survey Telescope. We thank the LSST Project for making their code available as free software at http://dm.lsst.org

The Pan-STARRS1 Surveys (PS1) have been made possible through contributions of the Institute for Astronomy; the University of Hawaii; the Pan-STARRS Project Office; the Max Planck Society and its participating institutes, the Max Planck Institute for Astronomy, Heidelberg, and the Max Planck Institute for Extraterrestrial Physics, Garching; the Johns Hopkins University; Durham University; the University of Edinburgh; Queen's University Belfast; the Harvard-Smithsonian Center for Astrophysics; the Las Cumbres Observatory Global Telescope Network Incorporated; the National Central University of Taiwan; the Space Telescope Science Institute; the National Aeronautics and Space Administration under grant No. NNX08AR22G issued through the Planetary Science Division of the NASA Science Mission Directorate; the National Science Foundation under grant No. AST-1238877; the University of Maryland; Eotvos Lorand University (ELTE); and the Los Alamos National Laboratory.

Based (in part) on data collected at the Subaru Telescope and retrieved from the HSC data archive system, which is operated by the Subaru Telescope and Astronomy Data Center at the National Astronomical Observatory of Japan.

## Data Availability

The data are publicly available at the HSC database (E. Jones et al. 2021).

## ORCID iDs

Swagata Biswas  https://orcid.org/0000-0001-7854-3230
Shubhrangshu Ghosh  https://orcid.org/0009-0008-0597-5348
Avyarthana Ghosh  https://orcid.org/0000-0002-7184-8004
Yogesh Wadadekar  https://orcid.org/0000-0002-1345-7371
Arijit Mukherjee  https://orcid.org/0000-0001-5052-4476
Shailesh Deshpande  https://orcid.org/0000-0001-8758-2557
Arpan Pal  https://orcid.org/0000-0001-9101-8051